**Title: Literature Review of Current Sustainability Assessment Frameworks and Approaches for Organizations**

**Primary Author:** Sarah Farahdel, PhD Student
**Co-Supervisors:** Chun Wang, Anjali Awasthi

# Abstract

This systematic literature review provides an overview of sustainability assessment frameworks (SAFs) developed by researchers in various industry-specific contexts. The review assesses sustainability indicator (SI) design approaches, including method of selection, relative importance evaluation, and interdependency assessment. The indicator selection process involves navigating complex criteria, and while diverse methods are used including, industry-specific literature reviews, stakeholder interviews, questionnaires, Pareto analysis, the SMART approach, and adherence to various sustainability standards and guidelines. Sustainability and performance indices offer dynamic assessments, with Fuzzy-AHP identified as one of the most robust techniques for evaluating the relative importance of SIs. Interdependency assessment methods include DEMATEL, VIKOR and its fuzzy applications, correlation analysis, causal models, and more. However, their static nature limits adaptability to dynamic organizational changes and sustainability priorities. Further strengths and limitations of the selected SAFs are presented, with a discussion on future research enhancements in this field. This review seeks to address the gaps the design approaches presented, contributing to the development of a comprehensive understanding of sustainability assessment frameworks for organizations and further provide valuable insights for policymakers, public administrators, organizational leaders, and researchers in enhancing sustainability practices. Future research is recommended for each of these design approaches, exploring multi-criteria decision-making models and hybrid approaches. Expanding sustainability evaluation within organizational levels or across supply chains is proposed for a holistic approach, emphasizing the need on adaptability to industry specifics and dynamic global adjustments, to further contribute to enhanced sustainability practices for organizations.

## 2.1   Introduction

The increasing recognition of sustainability as a critical factor for organizational success has led to the development of various Sustainability Assessment Frameworks (SAFs) designed for both public service sector organizations (PSOs) and private organizations (POs). This systematic literature review assesses existing selected SAFs, aiming to provide a comprehensive overview and assess their applicability in evaluating the sustainability performance of organizations. The paper examines multiple dimensions of SAFs, encompassing conceptual foundations, key components, evaluation criteria, methodologies, and practical implications. By critically analyzing the identified frameworks, the review seeks to unravel their strengths and limitations, offering insights into potential enhancements or modifications for improved effectiveness (Ahmad & Wong, 2018). Addressing significant gaps in the literature related to the selection method of criteria, sub-criteria, and indicators, as well as the evaluation of relative importance and interdependencies among sustainability indicators, this study contributes to a nuanced understanding of sustainability assessment frameworks. Furthermore, the research findings are set up to benefit policymakers, public administrators, organizational leaders, and researchers by providing valuable insights to enhance sustainability practices. The primary objective of this literature review is to is provide a focused analysis of some of the critical aspects and considerations in establishing a Sustainability Assessment Framework (SAF) for organizations. Research in this field has proven to have certain key limitations, in literature, for which this paper seeks to address through the following research objectives:

1. Investigate how the notion of 'sustainability' is strategically integrated into organizational frameworks.
2. Analyze the foundational criteria, sub-criteria and set of sustainability indicators (SIs) that make up for the structure of SAFs.
3. Examine the design approaches used in SAFs, specifically addressing:
    - The selection of sustainability indicators (SIs) within organizations.
    - The evaluation of the relative importance of each SI.
    - The interdependencies and impact of the selected SIs

These research objectives will help further understand the intricacies of SAFs, exploring commonalities, standardized approaches, and methodologies prevalent in both public and private sector organizations. Additionally, the study seeks to uncover potential advancements and progressions in the field of sustainability assessment, contributing to a deeper understanding of how organizations conceptualize and implement sustainability initiatives.

## 2.2 Methodology

A systematic literature review (SLR) was conducted for this research to explore existing sustainability assessment frameworks that have been developed in literature for different sectors and industries. For this review, a semi-structured snowballing approach was used. This approach aids in funneling down the search to pertinent references that are relevant to the objective of this study. Firstly, the research objective and question were identified, along with a selection of two main database sources: Scopus and Web of Science. Next, relevant associated keywords were identified in an iterative approach. The following keywords were used: "sustainability" assessment framework", "performance measurement", "organizations", "private organizations" and "public service sector organizations". A thorough mining was conducted of all the selected publications, including information such as titles, authors, year of publication, abstract, etc. Lastly, a systematic data synthesis of the selected articles was completed, through a thematic analysis of each article and its relevance to the defined research question. A comprehensive review of their titles, abstracts and texts was also done to only include articles that are most relevant to the topic of research. Then, a sufficiency test was done to assess if there were enough collected samples to contribute to the overall research question. **Figure 1** demonstrates the iterative phases taken to complete this systematic review.

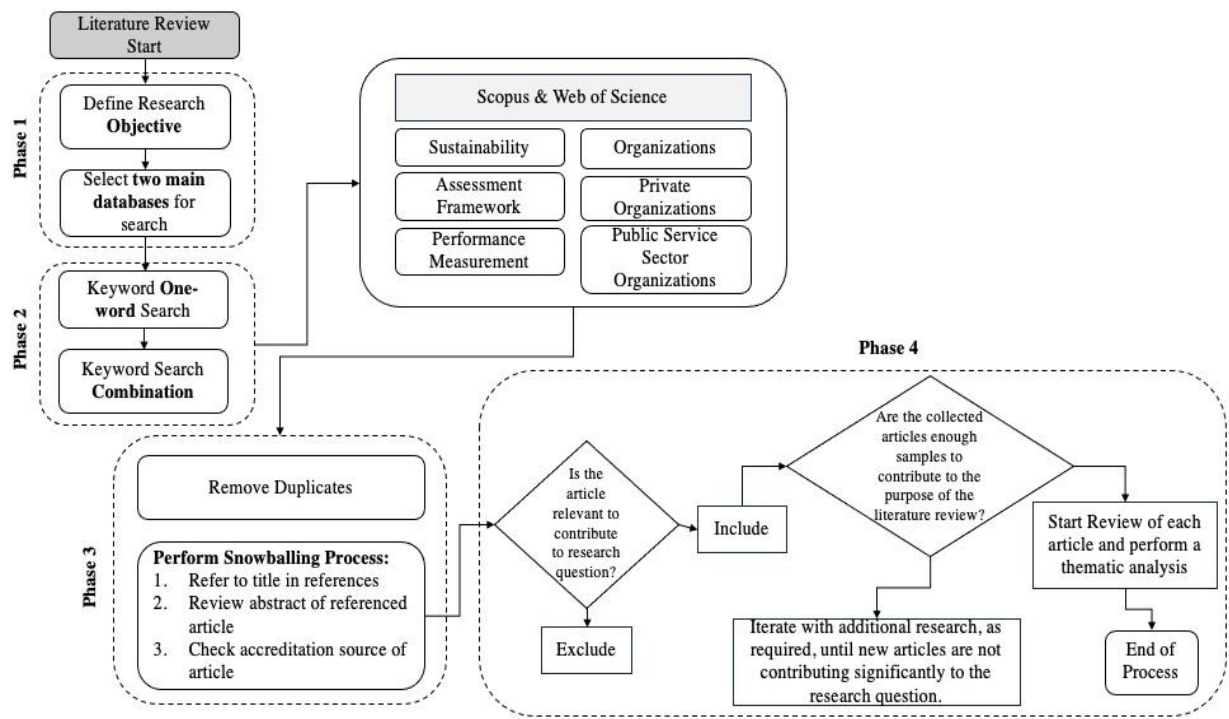

*Figure 1: Literature Review Process & Methodology*

The selection of these studies were limited to those that have introduced or presented sustainability assessment frameworks in their research and have omitted papers that only presented sustainability strategies or concepts in specific contexts. Many commonalities and differences have been found in each of these frameworks, which have been summarized in this study. A total of 55 sustainability assessment frameworks (SAFs) were selected and compared. A list of the assessed frameworks, with their associated sector or industry is described below. The earliest paper included in the dataset was published in 2002 and the latest was published in 2023. The analysis of these selected articles encompass a comparison on the basis of the following thematic criteria, sustainability integration method, tool structure, method of indicator selection, techniques used to evaluate the relative importance of indicators and to evaluate the interdependencies amongst indicators.

## 2.3. A Review of Existing Sustainability Assessment Frameworks (SAFs)

### 2.3.1 Sustainability Strategy Integration

#### *2.3.1.1  Sustainability Integration within Organizations*

Incorporating sustainability into corporate strategy and further into an effective sustainability assessment framework (SAF) has been a challenge for many organizations, as business strategies are to be deployed at different decisions levels of an organization (Medel-González et al., 2016). The integration of sustainability into an organization's evaluation process is narrow with significant gaps in both environmental and social impact areas (Da Ponte et al., 2020). Integrating sustainability as a concept into the decision-making process will enable the industry to overcome both present and emerging sustainability challenges (Husgafvel et al., 2015). It's important to consider both horizontal (including different perspectives or categories of impacts) and vertical integration (including different levels of decision making within an organization) when incorporation sustainability into an organizations strategy (Hacking & Guthrie, 2008; Lee, 2002). Recent studies have proven that sustainability assessment should go beyond just the organization as it has direct impacts to other organizations, key stakeholders and society, which brings about a new concept called Enterprise Sustainability Performance Systems Measurement (ESPMS) (Medel-González et al., 2016). A sustainability strategy is required to identify sustainable issues that are with internal and external environmental, economic, and social growth of an organization (Kaabi & Jowmer, 2018). Further, it's essential to create a link between sustainable development (SD) initiatives and broader organizational corporate goals, which should be considered as a dynamic concept that constantly evolves over time (Husgafvel et al., 2015; Hansen & Schaltegger, 2016). Based on the analysis of the selected SAFs in this study, it was observed that each one of these frameworks have a different level of incorporation of the three perspectives of sustainability; some with higher or lower emphasis on a particular dimension of the triple bottom line (economic, environmental and social). On average across all models the economic or 'profit' dimension had the least importance or presense within the frameworks, with environmental or 'planet' coming in second and social or 'people' coming in first, as the most important perspective considered.

#### *2.3.1.2  Sustainability Integration across Organizational Supply Chains*

This research demonstrates that the integration of sustainability is lacking across the organization's supply chain (SC) (Pagell & Wu, 2009), further requiring development in this domain (Gunasekaran et al., 2004; Ferreira & Silva, 2016). Sustainable development (SD) must be extended beyond organizational boundaries to incorporate the entire supply chain (Ferreira &

Silva, 2016). Reefke & Sundaram (2017) elaborates on the need to establish a clear and concise multi-objective SAF to fill this gap. Brandenburg et al. (2019) further expands on the need for more quantitative models to support a firm's strategic alignment with upstream and downstream SC partners. To evaluate the sustainability performance of the entire SC, an aggregation of the sustainability performance of each of the supply chain participants must be taken into account, to holistically look at the entire SC network (Saeed & Kersten, 2020). This however can be a challenging endeavour, requiring significant collaboration amongst stakeholders when measuring all three dimensions of sustainability across SCs (Wu et al., 2014). Additionally, the lack of data availability has also been considered a major concern (Ahi & Searcy, 2015), as it has been conventionally mainly focused on single firms as opposed to the entire SC (Hassini et al., 2012; Seuring, 2013). This limits the ability for leaders and key decision-makers to make rational decisions when not being able to see the impact of their decisions at all levels of the SC network. In summary, the incorporation of an effective sustainable strategy includes a comprehensive sustainability objective, as means to connect and integrate the goals and objectives of an organization both vertically and horizontally (Pei et al., 2010).

*Sustainability Integration Approaches*

An effective approach for incorporating sustainability into an organizations' strategy is to include comprehensive sustainability objectives, have a good connection between the existing goals and objectives of the organization, and have vertical and horizontal integration (Pei et al., 2010). Kaabi and Jowmer (2018) suggests having a planning team up front to contribute to the sustainability strategic plan, as strategic planning should depend on both financial and non-financial information. Ferreira and Silva (2016) demonstrated the importance of integrating the Balanced Scorecard (BSC) methodology, developed by Kaplan and Norton (2001) and made use of a strategy map, to facilitate the management of indicators and avoid introducing additional complexity. The strategy map is a useful tool that demonstrates the cause and effect relationship between the strategic objectives listed in each perspective and monitors the evolution of performance over time (Ferreira & Silva, 2016). The authors emphasized the need to involve multidisciplinary teams of internal stakeholders (including management, engineering, logistics, environmental, quality, purchasing and production departments) and external stakeholders (suppliers and non-governmental organizations). Hansen and Schaltegger (2016) presented an interesting typology of the generic SBSC architecture, demonstrating the hierarchy between the performance perspectives and the strategic objectives and how they relate to the overall corporate sustainability strategy. The authors suggests the leadership team needs to understand their corporate or institutional value system, which is the relationship between profit-making and sustainability. Then, based on their degree of proactivity preferred on corporate sustainability strategy, they can engage in the process of strategy formulation, specifying the degree of importance and relevance they want to give to each dimension (Hansen & Schaltegger, 2016). Kaabi and Jowmer (2018) further presented a four-stage implementation process to execute a sustainability strategy into a SBSC. The first step is translating the mission and vision to strategic objectives. Secondly, measuring the extend of which the objective is achieved using indicators and benchmarks. Thirdly, identifying sustainability resources dedicated to monitor the SBSC. Lastly, setting up the criteria to determine and compare the actual performance and efficiency of resources allocated to each objective. Although these all present great contributions, as approaches to integrating sustainability within an organizations' corporate strategy, further research is required to look into the impacts of existing and upcoming sustainability policies, governmental initiatives and guidelines to the sector or industry in which the organization is in.

### 2.3.2 Sustainability Assessment Framework (SAF) Model Structures

*Integration of Sustainability and Performance Measurement*

In the last two decades, research have increasingly developed sustainability assessment frameworks (SAFs) to quantifiably evaluate and assess their sustainability performance and progress towards their set goals. The concept of performance measurement (PM) combined with sustainability has been commonly used in recent literature as a foundation to build the model structure of these frameworks. The quantification of how a task is done efficiently and effectively can be referred to as performance measurement (PM) (Neely et al.,1995). Tajbakhsh & Hassini (2015) state that a PM model is necessary to maximize efficiency within business operations. Although there are numerous PM models that exists in literature, the Balanced Scorecard (BSC) developed by Kaplan and Norton in 1992, was the most cited by researchers and most widely used to expand upon through empirical studies and critical concepts (Hansen & Schaltegger, 2016; Na et al., 2020; Saroha et al., 2022). The traditional BSC used a 4-quadrant approach namely, financial, process, customer, and learning and growth, as an effective way to objectively measure performance, creating a link between strategy and day-to-day operations (Kaplan & Norton, 1992). The use of this model requires comprehensive participation of all stakeholders regarding operational performance (Tsai et al., 2020). Epstein and Wisner (2001) believe that to effectively determine what performance measures to use, the causal link between the actions taken and the impact of those actions on operational performance, customer value, sustainability performance, and financial performance are pivotal to consider. Figge et al. (2002) state that the BSC acts as a tool to identify the most strategically relevant aspects, using a top-down approach, and links them causally and hierarchically towards the long-term success of the organization. Researchers have since presented modified versions of the dynamic BSC model and offered distinctive methods in selecting and prioritizing indicators, incorporating different levels of sustainability and further used to this tool to collaborate amongst other supply chain (SC) partners (Voelpel et al., 2006). When measuring sustainability, there is need to balance economic, environmental, and social needs within the paradigm (Tajbakhsh & Hassini, 2015). Although the BSC has been used in management areas, the occurrence of the fourth industrial revolution that is transforming the environment and the strategic position of supply chains (SCs) have not been considered in its application (Saroha et al., 2022). The traditional BSC was later modified in 2002, by researchers of the German Ministry for Science and Education, to develop the sustainable balanced scorecard

(SBSC), which integrated two additional perspectives: environmental and social strategic objectives into its architecture (Hansen & Schaltegger, 2016). Scholars have highlighted the potential use of the SBSC as an effective tool to integrate conventional strategic management with corporate sustainability management, as it allows management to address goals in all three dimensions of sustainability in a single integrated system rather than requiring separate parallel systems (Figge et al., 2002). There are two ways sustainability can be incorporated in the BSC, one by incorporating sustainability objectives partly or fully into the four BSC perspectives or secondly it can be added as a dedicated perspective complimenting the four BSC existing perspectives (Epstein & Wisner, 2001; Figge et al., 2002). Some researchers critiqued this model and state that adding these two additional perspectives namely, social and environmental, will in fact shift the focus of stakeholders from strategy to PM, leading it more to a KPI scorecard rather than a strategy scorecard (Epstein & Wisner, 2001; Na et al., 2020a; Tsai et al., 2020). Other researchers have used the SBSC or the foundational elements of the model to create their own unique sustainable assessment framework (SAF) model. Hansen & Schaltegger (2016) state that sustainability modifications to the BSC framework are driven by instrumental, social, political, or normative theoretical perspectives. For this research, all frameworks including, SBSCs and other modified or unique models measuring sustainability will be referred to as SAFs.

*Comparative Analysis of Criteria used in existing SAFs.*

Further, a comparison was done looking at how each of these frameworks organized the sustainability indicators through the categorization of criteria. It's observed that a combination of the traditional BSC, modified sustainability BSC, the triple bottom line (TBL) and modified TBL perspectives were used to categorize the indicators selected for the framework. There were also some studies that included unique industry-specific frameworks. The distribution of the model structures used in these studies are illustrated in **Figure 2.**

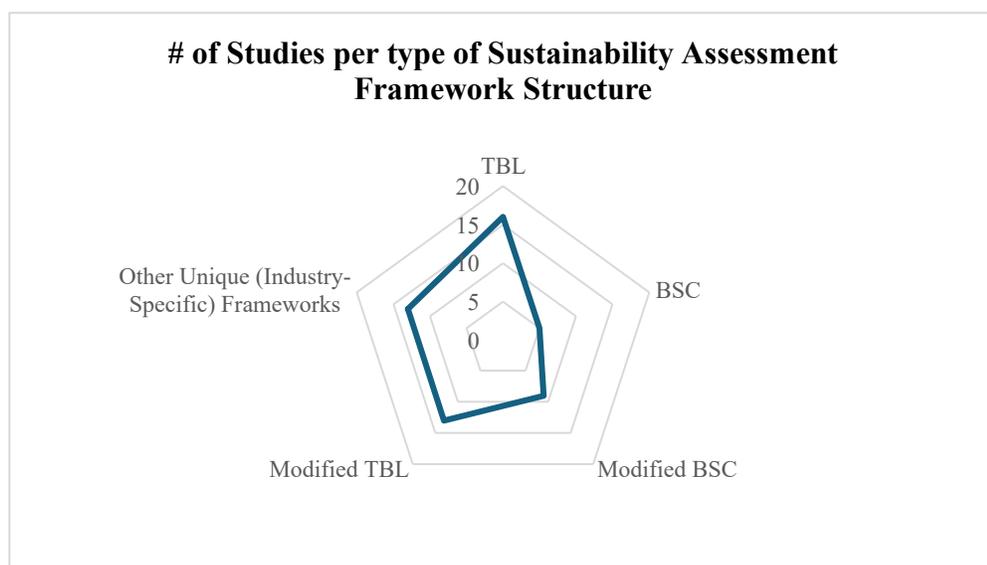

*Figure 1: Number of Studies per Sustainability Assessment Framework Structure*

It's evident that the TBL or modifications to the TBL perspectives have been the most popular use of a model structure, among other unique industry-specific frameworks. With the Balanced Scorecard (BSC) or modifications to the BSC being the least used amongst the frameworks analyzed. The differentiation used in all these models determines that there is not a standardized approach in assessing sustainability for a given organization and different perspectives or criteria are deemed important, based on the selected organization or specific industry. Further assessment would be required to evaluate how these perspectives can be clustered for a more unified and balanced model structure across all aspects of sustainability. Medel-González et al. (2013) presented a unique model where they used both the TBL and the BSC perspectives into their framework in a matrix form, derived from the MSSA 'Matrix of Sustainable Strategic Alignment' model developed by Oliveira et al. (2012), as illustrated in **Figure 4.** This model is claimed to provide true alignment between the sustainability strategy distributed over both the sustainability pillars and the SBSC criteria's (Medel-González et al.,2013), which can further be explored as a potential enhanced model structure.

| Pillars | Social | Economic | Environmental |
|---|---|---|---|
| Financial returns | Distribution profit to stakeholders (suppliers, distributors, communities, and other stakeholders) Create structures that supports other activities in the region | Maximization of profits Revenue maximization | Technology investments aligned to the concepts of clean productions and innovation Participation in sustainability indexes Participation in the program of carbon credits |
| Costumer's satisfaction & stakeholders interest's | Increased external perception about its social commitment through the development of social programs with public agencies or private | Increased participation in the market Customer retention Identifying new markets | Increased external perception about its environmental commitment with environmental programs |
| Internal processes | Transparent, ethical and fair treatment in intra-organizational relationships (selection, assessment and contact with all stakeholders) | Optimization of production processes internal and external Standardization establishment, reducing errors and waste | Demands of environmentally sound practices in intra-organizational processes Implementation of environmental standards |
| Learning & Growth | Cultural and educational development of the stakeholder process | Investment in the development of necessary and adherent competences to the organization's results | Understanding, development and multiplication of environmentally responsible culture |

*Figure 2: Matrix of Sustainable Strategic Alignment from Medel-González et al., 2013 derived from Oliveira et al., 2012*

## 2.3.3 Assessment of Design Approach

The design approach is then analyzed amongst these studies, following these five stages of analysis: indicator selection method, method of evaluating the relative importance of sustainability indicators and the assessment of interdependencies between the sustainability indicators.

### *2.3.3.1 Sustainability Indicator Selection Process*

Through identifying indicators and benchmarking, defined objectives can be measured more efficiently and can most often satisfy the primary needs of corporate management and global investors (Kaabi & Jowmer, 2018; Husgafvel et al., 2015). Developing the right indicators is a necessary step for any corporation or institution, as it will develop a sense of ownership of results, develop organizational learning and ensure the results truly reflect its values (Searcy et al. 2005). As there are many indicator sets that can be used, selecting the specific indicators, for different aspects of sustainability, that are right for an organization can a complicated endeavour (Husgafvel et al., 2015). Studies has shown that various researchers have used both quantitative and qualitative indicators to measure the performance of SC segments (Saroha et al., 2022).

- *Selection Criteria*

Certain studies have presented a set of criteria that an indicator should meet in order to be selected. The consolidation of criteria include ***target relevance or relevance to key objectives***, data availability, measurability and quality, validity, sensitivity, flexibility and responsive to change, accessibility and transparency, independent and standardized, unambiguous, reliability, widely used in international standard guidelines or policies, representability and comprehensibility, predictability, reasonable administrative burden / cost of collecting data, feasibility/operability, comparability, use of multi-criteria evaluation for all aspects of sustainability, consideration and alignment with stakeholder goals, , as presented in **Table 1**. Lima et al. (2014) add that data availability should be classified into short-term, medium-term and long-term and data quality to be classified as optimal, good and poor. In addition to this list, M. A. Adams & Ghaly, (2006) state that it's important to consider the need to present various indicators differently to different stakeholders. Furthermore, Singh et al. (2009) mention the importance of ensuring the indicator selected is clear, has a purpose, is flexible and adaptable to change and data is readily available to support the sustainment of the measure. The researchers provide a list of questions to consider when classifying a list of indicators namely the following:

1. What aspect of the sustainability does the indicator measure?
2. What are the techniques/methods employed for construction of index like quantitative/qualitative, subjective/objective, cardinal/ordinal, unidimensional/multidimensional.
3. Does the indicator compare the sustainability measure (a) across space ('cross-section') or time ('time-series') and (b) in an absolute or relative manner?
4. Does the indicator measure sustainability in terms of input ('means') or output ('ends')?

Selecting a set of indicators that meet all or even a few of the criterions listed above, can be a challenging task, hence it's important to recognize a sub-set of criteria's that matter and are realistic for the organization to meet.

| *Criteria* | *Reference* |
|---|---|
| **Target Relevance / Relevant to Key Objectives** | *Karjalainen & Juhola, 2019; Haghshenas & Vaziri, 2012; Alonso et al.; 2015; Yigitcanlar & Dur, 2010; Lundberg et al., 2009; Erol et al., 2011; Ramos et al., 2021* |
| **Data Availability, Measurability and Quality** | *Karjalainen & Juhola, 2019; Haghshenas & Vaziri, 2012; Alonso et al.; 2015; Lundberg et al., 2009; Yigitcanlar & Dur, 2010; Zito & Salvo, 2011; Erol et al., 2011; M. A. Adams & Ghaly, 2006 ; Singh et al. 2009* |
| **Validity** | *Alonso et al., 2015; Haghshenas & Vaziri, 2012; Karjalainen & Juhola, 2019; Yigitcanlar & Dur, 2010* |
| **Sensitivity, Flexibility and Responsive to change** | *Alonso et al., 2015; Haghshenas & Vaziri, 2012; Karjalainen & Juhola, 2019; Singh et al., 2009; Yigitcanlar & Dur, 2010* |
| **Accessibility and Transparency** | *Karjalainen & Juhola, 2019; Haghshenas & Vaziri, 2012; Alonso et al.; 2015; Zito & Salvo, 2011* |
| **Independent and standardized** | *Alonso et al., 2015; Haghshenas & Vaziri, 2012; Karjalainen & Juhola, 2019* |
| **Unambiguous** | *Alonso et al., 2015* |
| **Reliability** | *Alonso et al., 2015; Erol et al., 2011; Lundberg et al., 2009* |
| **Widely used in International Standard Guidelines or Policies** | *Lundberg et al., 2009; Yigitcanlar & Dur, 2010* |
| **Representability and Comprehensibility** | *Adams & Ghaly, 2006; Ramos et al., 2021; Yigitcanlar & Dur, 2010; Zito & Salvo, 2011* |
| **Predictability** | *Yigitcanlar and Dur, 2010* |
| **Reasonable Administrative Burden / Cost of Collecting Data** | *Erol et al., 2011* |

| **Feasibility/Operability** | *Ramos et al., 2021* |
|---|---|
| **Comparability** | *Zito & Salvo, 2011* |
| **Use of multi-criteria evaluation for all aspects of sustainability** | *M. A. Adams & Ghaly, 2006* |
| **Consideration and Alignment with Stakeholder Goals** | *M. A. Adams & Ghaly, 2006* |

*Table 1: List of Criteria for Sustainability Indicator Selection*

- *Methods of Selection*

Knowing where to select SIs from and how to select these indicators has been examined in literature. Different methods of indicators selection and/or a combination thereof have been used in literature including **industry specific literature review** (most commonly used approach) (Abbasi et al., 2023; Adewumi et al., 2023; Agarwal et al., 2022; Bhakar et al., 2018; Bhyan et al., 2023; Castillo & Pitfield, 2010; De Andrade Guerra et al., 2018; Erol et al., 2011; Gambelli et al., 2021; Gillis et al., 2015; Guo & Wu, 2022; Hansen & Schaltegger, 2016a; L. Karjalainen & Juhola, 2019; Leksono et al., 2018; Marletto & Mameli, 2012; Na et al., 2020b; Namavar et al., 2023; Neri et al., 2021; Pathak et al., 2021; Ramos et al., 2021; Saad et al., 2019; Saeed & Kersten, 2020; Sandhu et al., 2016; Saroha et al., 2022; Shen et al., 2013; Tafidis et al., 2017; Tsai et al., 2020; Wynder et al., 2013; Zito & Salvo, 2011), **stakeholder interviews and inputs** (M. A. Adams & Ghaly, 2006; Guo & Wu, 2022; Lim & Biswas, 2018; Lima et al., 2014; Marletto & Mameli, 2012; Olakitan Atanda, 2019; Pathak et al., 2021; Ramos et al., 2021, 2021; Sala et al., 2015), **questionnaires and surveys** (Adewumi et al., 2023; Lim & Biswas, 2018; Marletto & Mameli, 2012; Olakitan Atanda, 2019; Saad et al., 2019; Tsai et al., 2020), **pareto analysis** (Abbasi et al., 2023), **the SMART approach** (Gillis et al., 2015; Lundberg et al., 2009; Namavar et al., 2023), **related sustainability standards and guidelines** such as based on world-wide governance indicators (WGI), Global Reporting Initiative (GRI) Sustainability Reporting Standards, Down Jones Sustainability Index, ISO Standards (e.g. ISO 14031), Environmental Management System (EMS), etc.(M. A. Adams & Ghaly, 2006; Berardi, 2013; Bhyan et al., 2023; Ferreira & Silva, 2016; Husgafvel et al., 2015; Karjalainen & Juhola, 2019; Lundberg et al., 2009; Na et al., 2020; Olakitan Atanda, 2019; Saad et al., 2019; Saeed & Kersten, 2020; Shah et al., 2013; Stauropoulou & Sardianou, 2019; Tafidis et al., 2017; Zito & Salvo, 2011). Certain researchers have also presented their own unique methodological steps and other approaches to select sustainability indicators (SIs). The **Delphi Method** was another commonly used tool implemented by Olakitan Atanda (2019) and in fuzzy form by Namavar et al., (2023), and Tsai et al. (2020) to select indicators. This method was developed by Olaf Helmer and Norman Dalkey in the 1950's to use expert opinions through a feedback loop, allowing the participants to reassess their initial judgements on their selection and help resolve uncertainties regarding the preferences chosen (Tsai et al., 2020). The authors believe that this approach and combination will in fact reduce the number of interviewees required overall survey time and costs (Lee et al., 2018).

Although, there is not one universal method for selecting indicators, relying solely on individual approaches, such as existing literature reviews or stakeholder interviews, for sustainability indicator selection presents limitations. Focusing on literature reviews exclusively may disconnect chosen indicators from corporate objectives and industry standards. Similarly, depending only on stakeholder input or surveys may introduce biases and overlook essential quantitative metrics. Specific methods like the Delphi Method, Pareto analysis, or the SMART approach, if used in isolation, may not capture the full spectrum of qualitative and quantitative aspects, leading to imbalances in the selected SIs. While the Delphi Method facilitates expert opinions through a feedback loop, relying solely on it may lack the grounding in objective data and may not fully consider the organizational context or link to industry specific policies and guidelines. Pareto analysis, while effective in identifying vital few factors, might overlook less prioritized but still significant indicators. The SMART approach, emphasizing specific, measurable, achievable, relevant, and time-bound criteria, may not capture the full spectrum of qualitative aspects crucial for sustainability assessment. To enhance the effectiveness of SI selection, future research should adopt a holistic approach, integrating diverse methods, to achieve a more comprehensive and contextually relevant set of indicators aligned with organizational goals and specific industry standards. The exhaustive list of SI selection methods are illustrated in **Table 2**.

| # | Authors | Industry Specific Literature Review | Stakeholder Interviews/Inputs | Questionnaires or Surveys | Pareto Analysis | SMART | Related Sustainability Standards & Guidelines | Unique Methodological Steps | Others |
|---|---|---|---|---|---|---|---|---|---|
| 1 | Abbasi et al. (2023) | X | | | X | | | X | |
| 2 | Adewumi et al. (2023) | X | | X | | | | | Driver State Impact Response (DSIR) Approach |
| 3 | Bhyan et al. (2023) | X | | | | | X | | |
| 4 | Namavar et al. (2023) | X | | | X | | | X | Fuzzy Delphi Method and Sensitivity Analysis |
| 5 | Pan et al. (2023) | | | | | | | | Other Existing Infrastructure Sustainability Assessments (ISAs) |
| 6 | Azadi et al. (2023) | | | | | | | X | Directional Distance Function (DDF) and Network Data Envelopment Analysis (NDEA) model |
| 7 | Saroha et al. (2022) | X | | | | | | | |

| # | Author | | | | | | | Notes |
|---|---|---|---|---|---|---|---|---|
| 8 | Pathak et al. (2021) | X | X | | | | | |
| 9 | Ramos et al. (2021) | X | X | | | | | |
| 10 | Neri et al. (2021) | X | | | | | | |
| 11 | Gambelli et al. (2021) | X | | | | | | |
| 12 | Saeed & Kersten (2020) | X | | | | X | | |
| 13 | Na et al. (2020) | X | | | | X | X | Used techniques like text mining, sentiment mining, bag-of-words (BOW) to extract information |
| 14 | Tsai et al. (2020) | X | | X | | | | Fuzzy-Delphi method |
| 15 | Saad et al. (2019) | X | | X | | X | | |
| 16 | Stauropoulou & Sardianou (2019) | | | | | X | | |
| 17 | Karjalainen and Juhola | X | | | | X | | |
| 18 | Lim and Biswas (2018) | | X | X | | | | |
| 19 | Leksono et al. (2018) | X | | | | | | |
| 20 | Kaabi and Jowmer (2018) | | | | | | X | 5M method used to recruit resources to part-take in indicator selection (Man, Materials, Methods, Machine, Money) |
| 21 | de Andrade Guerra et al. (2018) | X | | | | | | |
| 22 | Bui et al. (2017) | | | | | | X | |
| 23 | Arodudu et al. (2017) | | | | | | X | Basic Elements Table used (Where, What, Who, When?) and LCA and LCT Methodologies used |
| 24 | Tafidis, Sdoukopoulos, and Pitsiava-Latinopoulou (2017) | X | | | | X | X | |
| 25 | Ferreira and Silva | | | | | X | X | |
| 26 | Hansen and Schaltegger (2016) | X | | | | | | |
| 27 | Gillis et al. (2016) | X | | | X | | | |
| 28 | Sala et al. (2015) | | X | | | | | |
| 29 | Husgafvel et al. (2015) | | | | | X | | |
| 30 | Alonso et al. (2015) | | | | | | X | |
| 31 | Palma Lima et al. (2014) | | X | | | | | |
| 32 | Shi & Lai (2013) | | | | | | X | |
| 33 | Berardi (2013) | | | | | X | | |
| 34 | Wynder et al. (2013) | X | | | | | | |
| 35 | Shah et al. (2013) | | | | | X | | |
| 36 | Toth-Szabo and Varhelyi (2012) | | | | | | X | |
| 37 | Waheed et al. (2011) | | | | | | X | Driving pressure-state-exposure-effect-action (DPSEE) linkage-based approach |
| 38 | Ramani et al. (2011) | | | | | | X | |
| 39 | Zito and Salvo (2011) | X | | | | | X | |
| 40 | Marletto and Mameli (2010) | X | X | X | | | X | |
| 41 | Yigitcanlar & Dur (2010) | | | | | | | SILENT Model |
| 42 | Castillo and Pitfield (2010) | X | | | | | X | |
| 43 | Lundberg et al. (2009) | | | | X | X | | |
| 44 | Adams and Ghaly (2006) | | X | | | X | X | |
| 45 | Figge et al. (2002) | | | | | | X | |
| 46 | Sandhu et al. (2016) | X | | | | | | |
| 47 | Agarwal et al. (2022) | X | | | | | | Best-Worst Method (BWM) |
| 48 | Whitehead (2017) | | | | | | X | Materiality Analysis |
| 49 | Bhakar et al. (2018) | X | | | | | | |
| 50 | Atanda (2019) | | X | X | | | X | Delphi Method |
| 51 | Guo and Wu (2022) | X | X | | | | | |
| 52 | Shen et al. (2013) | X | | | | | | |
| 53 | Erol et al. (2011) | X | | | | | | |
| 54 | Ramos et al. (2021) | | X | | | | | |
| 55 | Medel-Gonzalez et al. (2013) | | | | | | X | GRI guidelines and ISO 14031 |

*Table 2: List of Methods used to Select the appropriate set of Sustainability Indicators (SIs)*

- **Sustainability Indices**

In addition to traditional sustainability indicators, Karjalainen & Juhola (2019) state that performance indices, including composite sustainability indices (CSIs) have been used, as well as modelling and simulations to assess various public transportation infrastructure projects and measure their sustainability performance in a more dynamic form. Zito and Salvo (2011), Husgafvel et al. (2015), Medel-González et al. (2016) and Stauropoulou & Sardianou (2019) used Sustainability Indices (SINDEX) to assess sustainability performance in their studies. Indexing is a method used by aggregating a group of indicators into a single value (Zito & Salvo, 2011), to provide a simplified representation of performance and make it easier for key decision-makers to grasp the overall trends or performance in a certain area of the business. The benefit to using indices over a set of indicators is that it allows for easier benchmarking and comparison between various entities, industries or regions, while also providing a standardized approach to allow for easy comparison across different organizations and supply chains including both upstream and downstream activities or sectors (Husgafvel et al., 2015). Indices also provide a great means to align with broader organizational strategic goals and objectives, enabling decision-makers to focus on overarching sustainability priorities. One of the shortcomings of using sustainability indices can be an over-simplification of the complex nature of sustainability, which can result in a loss of specific details required to capture the challenges or true improvement of an organization in a given area.

## 2.3.3.2 Evaluating Relative Importance of Sustainability Indicators (SIs)

Evaluating the relative importance or 'weighting' of sustainability indicators (SIs) is crucial for improved prioritization, resource allocation, increased accountability, reduced subjectivity, and effective long-term planning (Bhyan et al., 2023). The establishment of a robust Sustainable Assessment Framework (SAF) weighting mechanism presents a significant challenge for stakeholders (Bhyan et al., 2023). Various methodologies are employed for this purpose, each offering unique advantages and limitations. The table below summarizes different Multi-Criteria Decision-Making (MCDM) models and techniques utilized in selected studies. Researchers have recognized the uncertainties present in sustainability assessments, are crucial for policymakers deciding on risk acceptance or action plans to meet set targets (Sala et al., 2015).

After selecting the right indicators for the scorecard, the next step would be to assess whether they all contribute the same value to the overall strategic objective. Weighting helps define the level of impact of one indicator to the overall performance of a set perspective. This next section summarizes the different approaches taken by researchers to this end, additionally describing the purpose, advantage and limitations of each method used. **Table 2** presents a list of multi-criteria decision-making (MCDM) models and techniques that were used in the selected studies, including the names of the original creators of such models. Note several authors have used the fuzzy application of these models in their studies to account for the uncertainty in the stakeholder's input.

From this analysis, the most common techniques presented to evaluate the relative importance of SIs were average weights, normalization or scoring approach, the Analytical Hierarchical Process (AHP), fuzzy-AHP, and the aggregation method, in respective order of application. Other distinct approaches were also presented, including fuzzy TOPSIS, fuzzy entropy, Additive Ratio Assessment (ARAS), Driver-Focused Prioritization (DFP) and the the Importance-Performance Analysis (IPA) model. The remainder of the studies did not present an approach to evaluate the relative importance of the SIs and hence considered them all as equal importance.

| Method | # of Studies | References |
|---|---|---|
| **AHP** | 5 | *Abbasi et al. (2023), Adewumi et al. (2023), Stauropoulou & Sardianou (2019), Castillo and Pitfield (2010), Atanda (2019)* |
| **Fuzzy AHP (f-AHP)** | 3 | *Bhyan et al. (2023), Bui et al. (2017), Guo and Wu (2022)* |
| **Normalization or Scoring Approach** | 5 | *Namavar et al. (2023), Shi & Lai (2013), Ramani et al. (2011), Castillo and Pitfield (2010), Ramos et al. (2021)* |

| **Average Weight** | 9 | *Pan et al. (2023), Lim and Biswas (2018), Leksono et al. (2018), Alonso et al. (2015), Palma Lima et al. (2014), Wynder et al. (2013), Shah et al. (2013), Waheed et al. (2011), Zito and Salvo (2011)* |
|---|---|---|
| **Weighted SUM** | 2 | *Gillis et al. (2016), Berardi (2013)* |
| **Aggregation Method** | 3 | *Abbasi et al. (2023), Palma Lima et al. (2014), Marletto and Mameli (2010)* |
| **Fuzzy TOPSIS (*f-TOPSIS*)** | 1 | *Shen et al. (2013)* |
| **Fuzzy Entropy** | 1 | *Erol et al. (2011)* |
| **Additive Ratio Assessment (ARAS)** | 1 | *Agarwal et al. (2022)* |
| **Driver-Focused Prioritization (DFP)** | 1 | *Whitehead (2017)* |
| **Importance-Performance Analysis (IPA) Model** | 1 | *Gambelli et al. (2021)* |

*Table 2: List of techniques used to evaluate relative importance of the selected indicators*

The average weight technique used by researchers in this case, calculates the mean value of weights assigned to each SI by a set of selected participants part-taking in the process (Lim & Biswas, 2018), providing a simple score (Leksono et al., 2018). This approach lacks however in addressing any sensitivity to variations and is subject to stakeholder bias, leading to a less nuanced understanding of the true significance of each SI within the overall assessment framework. Conversely, the weighted sum technique used by Gillis et al. (2016) and Berardi (2013), assigns weights to each indicator and aggregates them to calculate an overall score, which leads to similar subjectivity in assigning weights, potential biases introduced by disproportionate emphasis on certain SIs, and lacks in a comprehensive and unbiased representation of sustainability performance, similar to the limitations present in using the aggregation method. To address this limitation, the Analytical Hierarchical Process (AHP) developed by Saaty (1980) selects the hierarchy based on weighted aggregation and pair-wise comparison, helping to prioritize the SIs based on their relative importance to the overall decision objectives (Saaty, 1980). The reliability of traditional AHP however can be compromised due to uncertainty, as it uses a 9-point ranking scale; to overcome this constraint, Fuzzy AHP (f-AHP) can be applied to handle uncertainties effectively (Bhyan et al., 2023). Fuzzy sets, derived from the fuzzy theory developed by Zadeh in 1965 have been used to digest the vagueness or incompleteness of information, eliminating the impact of diverse human subjectiveness or consciousness. FAHP offers a notable advantage over other MCDM techniques by providing reliable and statistically robust results with a smaller sample size; approximately 5-10 responses from experts can deliver acceptable weights (Bhyan et al., 2023; Kaganski et al., 2018).

As an alternative approach, researchers Shen et al. (2013) used fuzzy TOPSIS (Technique for Order of Preference by Similarity to Ideal Solution) method, originally developed by researchers by Yoon and Hwang in 1981, which is based on the theory that the selected alternative solution should have the shortest distance from the positive ideal solution (PIS) and the farthest distance from the negative ideal solution (NIS) (Shih et al., 2007). Although this technique is having an ability to handle uncertainties and vagueness in decision-making processes, offering a more flexible and nuanced approach to capturing imprecise information, it still however has susceptibility to the choice of linguistic terms and functions, which can further cause subjectivity.

Unique approaches presented by Agarwal et al. (2022) (ARAS), Whitehead (2017) (DFP) and Gambelli et al. (2021) (IPA), all suffer from significant time investment and manual effort required from selected participants and moreover they still hinder from subjectivity and bias approaches. Gambelli et al.'s (2021) Importance-Performance Analysis (IPA) model linked to the BSC model, classifying each business performance indicator (BPI) into four categories, which are usually defined as: 'Keep up', 'Concentrate here', 'Possible overkill' and 'Low priority', which is a very familiar approach used by industry leaders in practice.

The FAHP method appears as the most robust approach in the examined studies for evaluating the relative importance of the SIs, due to its ability to handle uncertainties, reduce subjectivity and digest vague information. Future research should explore integrating alternative techniques with FAHP, to allow for allow for dynamic adjustments, as required. Notably, the identified techniques demonstrated a static nature, unable to accommodate dynamic parameter changes or evolving priorities over time. This limitation demands manual repetition of evaluations from key participants and stakeholders, potentially hindering the adaptability of the sustainability assessment framework.

*2.3.3.3 Interdependency Assessment*

Researchers have utilized distinct methodologies to explore interdependencies among sustainability indicators, as presented in **Table 3**.

| *Method* | *References* |
| --- | --- |
| *DEMATEL and ANP* | *Leksono et al. (2018)* |
| *Fuzzy DEMATEL* | *Tsai et al. (2020)* |
| *Fuzzy DEMATEL and VIKOR* | *Guo and Wu (2022)* |
| *Correlation Analysis and Cluster Analysis* | *Alonso et al. (2015)* |
| *Causal Models (BSC and Strategy Map)* | *Wynder et al. (2013), Waheed et al. (2011), Figge et al. (2002)* |
| *Fuzzy Multi-Attribute Utility (F-MAUT)* | *Erol et al. (2011)* |

*Table 3: List of MCDMs and techniques used to assess the interdependencies amongst the selected indicators*

Leksono et al., (2018) employed the Decision-Making Trial and Evaluation Laboratory (DEMATEL) and Analytical Network Process (ANP) to delineate relationships between perspectives and indicators. DEMATEL, a multiple-attribute decision-making (MADM) method developed by the Battelle Memorial Institute (Guo & Wu, 2022), elucidates direct or indirect relationships through matrix analysis (Leksono et al., 2018), while ANP, an advanced form of Analytical Hierarchy Process (AHP), considers interdependencies between variables. Although, the researchers successfully designed relationships between perspectives and indicators, there lies potential subjectivity during pairwise comparisons and the need for expertise in assigning weights to establish interdependencies while using these techniques. To address the incompleteness of data or information available, the fuzzy Decision-Making Trial and Evaluation Laboratory (f-DEMATEL) technique was used by (Tsai et al., 2020). Lin and Tzeng (2009) confirm DEMATEL's reliability for recognizing interdependencies among indicators, aiding in understanding driven and influencing indicators by visualizing dependencies through a visual network. Although, potential challenges in selecting appropriate linguistic terms and membership functions, can result from this technique leading to subjectivity and reduced robustness. Guo and Wu (2022) extended this research, utilizing a fuzzy hybrid method incorporating AHP, DEMATEL, and VIKOR to evaluate social sustainability in supply chains (SCs). The VIKOR VIekriterijumsko KOmpromisno Rangiranje) is a Serbian technique addresses conflicting criteria through multi-criteria optimization and compromise solutions (Opricovic & Tzeng, 2007). Although this proves to be a novel approach, challenges can arise in the potential complexity of combining these methods, requiring thorough consideration and expertise.

The correlation analysis and cluster analysis presented by (Alonso et al., 2015), provides valuable insights into indicator interdependencies, but can overlook more intricate relationships and dependencies within a more complex system or setting. While causal models like the Balanced Scorecard (BSC) and Strategy Map used by Wynder et al. (2013), Waheed et al. (2011), Figge et al. (2002) offer a comprehensive view, their limitation lies in static representations, lacking the ability to adapt to dynamic changes over time. Erol et al. (2011) conversely used the fuzzy entropy and fuzzy multi-attribute utility (F-MAUT) framework to quantitatively assesses sustainability across supply chains. The F-MAUT theory, originally founded by Hwang and Yoon (1981), combines scoring techniques and optimization models, facilitating better decision-making and stakeholder engagement. The MAUT theory is developed to help decision makers assign utility values to devices in terms of single attribute utility functions and combine individual evaluations to obtain overall utility values (Erol et al., 2011) based on the decision maker's preferences (Hwang and Yoon, 1981). Although this method has proven to be beneficial in demonstrating sustainability progress and improved decision-making, it does however rely on qualitative data collection methods, introducing perception-based biases while not fully capturing quantitative measures.

**Table 4** below demonstrates a consolidation of the benefits and disadvantages of each of these methods used for the purpose of determining the interdependencies amongst sustainability indicators, sub-criteria, and criteria.

| Method | Benefits | Disadvantages |
|---|---|---|
| *DEMATEL & ANP* | - Provides systematic approach for analyzing interdependencies amongst indicators<br>- Considers direct and indirect relationships<br>- Allows prioritization of variables based on influence identified | - Sensitive to changes in input data and highly subject to biases<br>- Heavily reliable on expert opinion |
| *Fuzzy DEMATEL* | - Incorporates fuzzy logic to handle uncertainties and vagueness of data<br>- Allows for a realistic representation of SIs | - Requires additional effort to determine fuzzy functions<br>- Interpretation of fuzzy results can be challenging for decision-maker |
| *Fuzzy DEMATEL & VIKOR* | - Combines strengths of both fuzzy DEMATEL and fuzzy VIKOR for a more robust analysis<br>- Incorporates fuzzy logic | - Increased complexity in implementation and perhaps interpretation<br>- Requires higher level of expertise in methodologies |
| *Correlation Analysis and Cluster Analysis* | - Provides valuable insights on relationship amongst variables<br>- Straightforward and basic approach | - Assumes linear relationships amongst variables<br>- Cannot accurately capture non-linear or complex relationships |
| *Causal Models* | - Presents a strategic perspective on SIs, allowing decision-maker to understand cause-and-effect relationships | - Requires well-defined comprehensive set of SIs upfront<br>- Dependent on quality of initial strategy formulation<br>- Limited ability to quantitatively capture complex relationships |
| *Fuzzy MAUT* | - Integrates fuzzy logic with utility theory<br>- Allows decision-makers preferences to be considered | - Computationally intensive for large-scale problems<br>- Requires careful collection of preferences and fuzzy sets<br>- Heavily reliable on decision-makers preferences |

Although each technique provides valuable insights, it's important to acknowledge their respective limitations. Future research directions should aim to enhance these tools by introducing adaptability to changing parameters and consider the potential integration of artificial intelligence techniques for improved dynamism in sustainability assessments.

## 2.4  Key Challenges in Sustainability Assessment Frameworks

Various challenges stem from establishing a Sustainability Assessment Framework (SAF) for organizations, whether in the public or private sector. Key aspects include sustainability strategic objectives integration, stakeholder engagement, data collection and reporting, supply chain considerations, continuous improvement, technology, innovation, legal and regulatory compliance, and cultural and organizational change. Data collection can be a complex and time-consuming task, especially if primarily constrained by data availability (Ahi & Searcy, 2015). Customer and supplier inputs, especially when measuring cross-functionally, pose challenges in terms of complexity (Hervani et al., 2005). Challenges in data collection contain issues such as data availability, data quality, bias, ownership, and methods (Ferreira & Silva, 2016). Researchers, including Husgafvel et al.'s , (2015), Leksono et al. (2018) and Gambelli et al. (2021) relied on questionnaires or surveys, leading to potential time inefficiencies, subjectivity, and biased information. The need for a more systematic approach, leveraging reliable and accessible data sources, is crucial to prevent conflicts and confusion among stakeholders. While there's significant research on sustainability within corporations, evaluating sustainability across supply chains (SCs) is limited, with a particular need for integration of both vertical and horizontal aspects (Ferreira & Silva, 2016; Gunasekaran et al., 2004). There is a need for organizations to collect data from their organizational supply chain as well, rather than focusing it on their single firm (Ahi & Searcy, 2015; Hassini et al., 2012; Seuring, 2013). Sustainable Supply Chain Management (SSCM) emphasizes strategic, transparent integration, hence there is a need to integrate social and environmental considerations within the SC context (Brandenburg et al., 2019; Carter & Rogers, 2008; Pagell & Wu, 2009; Reefke & Sundaram, 2017; Zhu & Sarkis, 2006). Although, not mentioned in most of these studies, continuous improvement remains vital step for organizational success once the SAF has been set up, and incorporating models like the Plan-Do-Check-Act (PDCA) can be an option (Ferreira et al., 2016). As sustaining efforts and progress in SAFs requires continuous monitoring and updating (Sala et al., 2015). Future research directions should address these challenges by focusing on systematic and reliable data collection methods, integrating continuous improvement models, and expanding the evaluation of sustainability across entire supply chains, considering both social and environmental dimensions.

## 2.5  Discussion and Future Research

The review assessed the design approach of SAFs, including the SI selection method, relative importance evaluation, and interdependency assessment. The SI selection process requires navigating complex criteria, including target relevance, data availability, and validity. While various methods like literature reviews, stakeholder interviews, and questionnaires are employed for selection, where solely relying on individual approaches may pose limitations. Furthermore, sustainability and performance indices offer a dynamic assessment, simplifying representation for decision-makers. Fuzzy AHP emerges as a robust technique amongst all presented for evaluating the relative importance, while interdependency assessment methods encompass DEMATEL, fuzzy DEMATEL, correlation analysis, causal models, and others. While the methods presented by researchers offer unique contributions to the field, there still lies drawbacks from their static nature, limiting their ability to adjust seamlessly to the dynamic changes of the organization and their sustainability priorities.

Future research in field could focus on evaluating the efficiency of indicator selection methods proposed as well as those presented for evaluating the relative importance and interdependencies of indicators, prompting a deeper exploration of multi-criteria decision-making models and potential hybrid approaches to enhance accuracy and an improved output. A systematic review expanding sustainability evaluation within hierarchical levels of an organization or across their supply chain, can offer a more holistic approach to sustainability assessment. Additionally, comparative analyses across diverse geographical or economic settings may uncover contextual drivers influencing sustainability and assessment and reporting practices. Future research in this field should prioritize a nuanced understanding of sustainability assessment for organizations, emphasizing their adaptability to industry specifics and global adjustments and fluctuations over time.